\documentclass[aps,twocolumn,showpacs,epsfig]{revtex4}
\usepackage{graphicx}
\usepackage{epsfig}
\usepackage{times}
\newcommand{\be}{\begin{equation}}
\newcommand{\ee}{\end{equation}}
\newcommand{\bea}{\begin{eqnarray}}
\newcommand{\eea}{\end{eqnarray}}

\begin{document}
\title{Force induced unfolding of bio-polymers in a cellular 
environment: A model study}
\author{Amit Raj Singh$^1$, Debaprasad Giri$^2$, 
and Sanjay Kumar$^1$}
\affiliation{$^1$Department of Physics, Banaras Hindu University,
Varanasi 221 005, India \\
$^2$Department of Applied Physics, IT, Banaras Hindu University,
Varanasi 221 005, India 
}
\date{\today}
                                                                                                
\begin{abstract}
Effect of molecular crowding and confinement experienced by 
protein in the cell during unfolding has
been studied by modeling a linear polymer chain on a 
percolation cluster. It is known that internal structure 
of the cell changes in time, however, they do not change 
significantly from their initial structure. In order 
to model this we introduce the correlation among the 
different disorder realizations. It was shown that the 
force-extension behavior for correlated disorder in 
both constant force ensemble (CFE) and constant distance 
ensemble (CDE) is significantly different than the one 
obtained in absence of molecular crowding. 
\end{abstract}
\maketitle

\section{Introduction}

Understanding of the structure and function of bio-polymers in vivo
by analyzing it in vitro is one approach \cite{rief2,busta3,itzh,lemak}, 
but another route is to perform the analysis in presence of the environment 
similar to in vivo \cite{ellis,matou,goodsell,mueller}. It is now known that 
the interior of the cell contains different kind of biomolecules 
like sugar, nucleic acids, lipids etc. These macromolecules 
occupy about $40\%$ of the total volume with steric repulsion 
among themselves. This confined environment induces  phenomena 
like ``molecular confinement" and ``molecular crowding" and 
has major thermodynamic and kinetic consequences on the cellular 
processes \cite{minton,cheung,np2009}. 
In recent years, Single Molecule Force Spectroscopy (SMFS) 
experiments were mainly performed in vitro (absence of 
molecular crowding) to understand the cellular processes 
\cite{rief2,rief1,busta1,busta2}. 
Theoretical modeling of such processes with simplified 
interactions 
\cite{fixman,degennes,doi,vander,li0,somen1,maren1,maren2,kumar,zhou} 
amenable to statistical mechanics have been used extensively 
to compare the outcomes of these experiments. Such models 
despite their simplicity, have been proved to be quite predictive 
and provided many important information about the 
cellular processes.  In all such modeling, the confinement 
imposed by the cellular environment has been ignored and 
surrounding environment has been considered as homogeneous. 
For example in protein unfolding each amino acid (monomer) 
interacts with its fixed number (depending on the lattice) 
of nearest neighbors \cite{li0,somen1,maren1,maren2,kumar,zhou}.

\begin{figure}[t]
\includegraphics[width=2.5in]{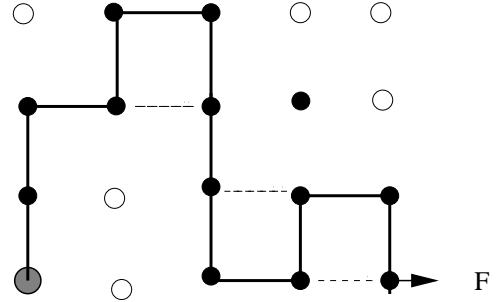}
\caption{Schematic representation of a polymer chain on a disordered lattice.
The filled circle represents the available site and open circle represents 
the occupied sites.  The imposed restriction of $p_c$ above than the 
value 0.5928
represents the unavailability of fraction of sites (``volume exclusion" 
which is about $40 \% $ as seen in the cell). One end of the polymer 
chain is kept fixed while a force $F$ has been applied at the other end.
}
\label{fig-1}
\end{figure}

In vivo, cellular environment and interactions involved in the 
stability of protein remain no more homogeneous. Hence in 
theoretical modeling, it is essential to use the underlying 
lattice to be disorder (or  random) in order to study the effects 
of the cellular environment. If different biomolecules do 
not move with time the effect of such confinement is termed 
as ``molecular confinement".  If they move, then effect is 
termed as "molecular crowding". From the statistical mechanics 
perspective for the ``molecular confinement", quenched averaging 
is appropriate while for the ``molecular crowding" constrained annealed 
averaging will be suitable. It is pertinent to mention here 
that  unfolding process is a dynamical phenomenon of the 
order of micro seconds to  few seconds. In such a short 
time span, internal structure of cell does not access 
all possible conformations. Hence, experimental situation 
does not require the averaging over all possible internal 
conformations ($\rightarrow \infty$) of the the cell. 
Therefore, in present study, we restrict ourselves to finite, but a set of 
large disordered realizations which roughly reproduce 
the effect of the internal structure of the cell.

The aim of this paper is to study the effect of an applied force
on a polymer trapped in a random environment as shown in Fig. 1.
The random environment mimics the effect of crowding agents 
as seen in the cell.  The confinement of the polymer 
to a restricted portion of phase space leads to the loss in entropy. 
Therefore, effect of the applied force on the reaction co-ordinate (in this 
case end to end distance) is mainly determined by the loss of 
configurational entropy and gain in internal energy because of  the 
nearest neighbor interaction ($\epsilon$). Because of the 
surrounding environment the average number of interactions per monomer 
is not constant, but may vary in between  0 to 2. This introduces 
heterogeneity in the interaction along the chain even in case of  
homopolymers.
It is important to mention here that in this case the precise 
information about the reaction co-ordinate, probability distribution of 
the reaction co-ordinate on the parameters of the system is difficult 
to obtain analytically, therefore, one has to resort on numerical treatments. 

\begin{figure*}[t]
\includegraphics[width=5in]{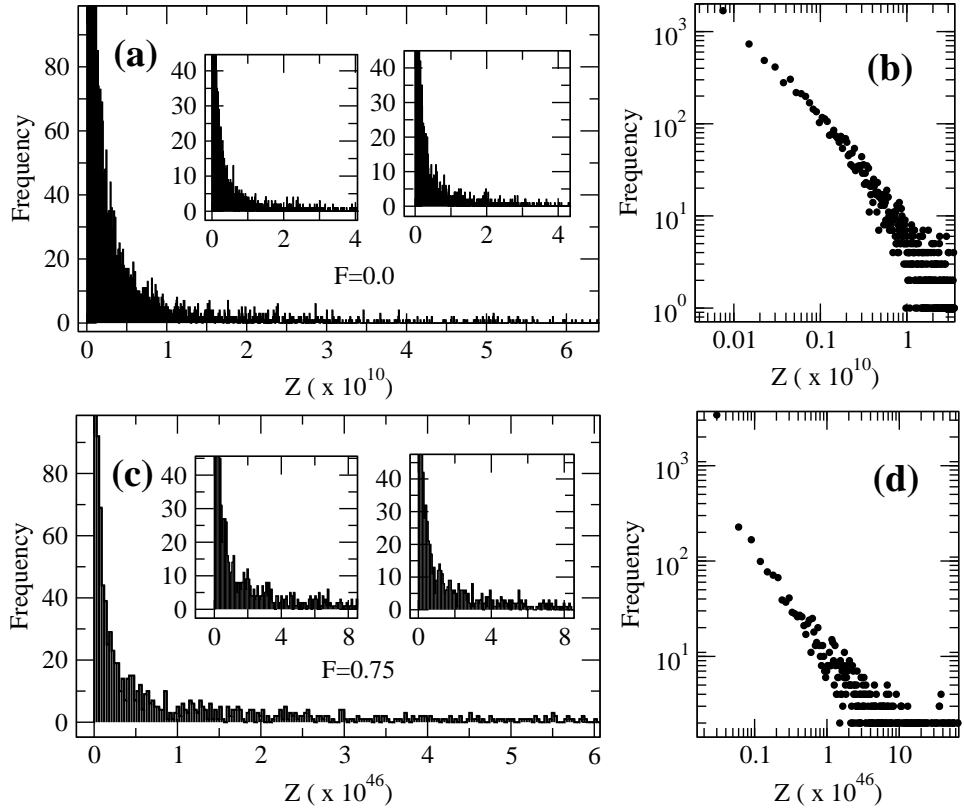}
\caption{(a) Histogram of $Z$ at $F=0$ and $T = \infty$; (b) Corresponding
histogram in log-log scale shows the power law behavior;
(c \& d) Same as (a, b) but for $F=.75$ and $T = 0.3$.}
\label{fig-2}
\end{figure*}

The paper is organized in the following way. In sec. II, we introduce the 
model to mimic the cellular environment and describe how we are considering the 
random correlated disorder environment. We also briefly discuss the 
exact enumeration 
technique in this section. In sec. III we present our main results 
obtained in constant force ensemble (CFE) and constant distance 
ensemble (CDE). We show the force-extension behavior in presence of 
crowding agents significantly different than the one obtained for the 
pure case which represents the unfolding process in vitro. 
We briefly discuss the consequence of molecular crowding and confinement 
in sec. IV.

\section{Model and Method}

\subsection{Random Environment: Correlated Disorder}

We model a polymer chain described by the self attracting self 
avoiding walks (SASAWs) \cite{degennes} confined in a random environment. 
Such environment can be obtained by using the method developed 
for the percolation cluster. We choose probability for an 
unoccupied site equal to $p$ which is just above the percolation 
threshold $p_c$ (= 0.5928) \cite{bkc,bkc1,janke,stauffer}. 
This will offer a wide class of the distribution of clusters. Cluster
size can vary from the isolated site to the length of the 
system. For the collapsed state the ground state energy arising 
due to the nearest neighbor  interaction will not be constant.
It may be noted that in case of regular lattice, ground 
state energy is $-\epsilon N$ which is a constant.

\begin{figure}[t]
\vspace{.5in}
\includegraphics[width=3in]{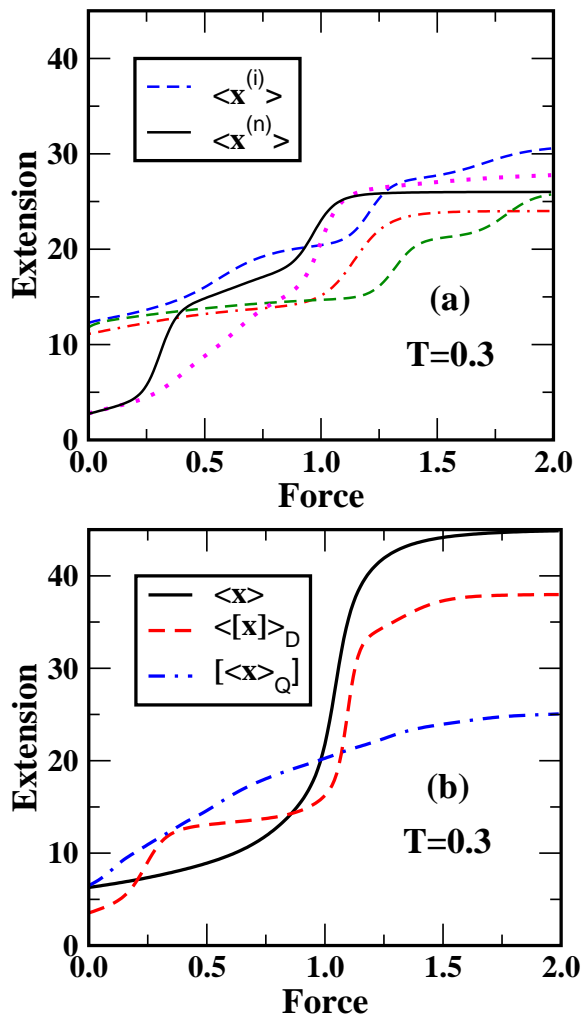}
\caption{Figures show the variation of reaction co-ordinate with force
at different temperature in CFE: (a) Representative
F-x curve for different disorder realizations at T=0.3. $<x^{(n)}>$ is
for initial (native) realization and $<x^{(i)}>$ is for other realizations
which differ from native by $10\%$; (b) F-x curve
for approximate annealed average and sample averaged quenched disorder
at temperature (T=0.3);
The native (initial realization), approximate annealed case and
individual quenched disorder
show the multi step plateaus which are absent in pure case.}
\label{fig-3}
\end{figure}

\begin{figure*}[t]
\vspace{.5in}
\includegraphics[width=5in]{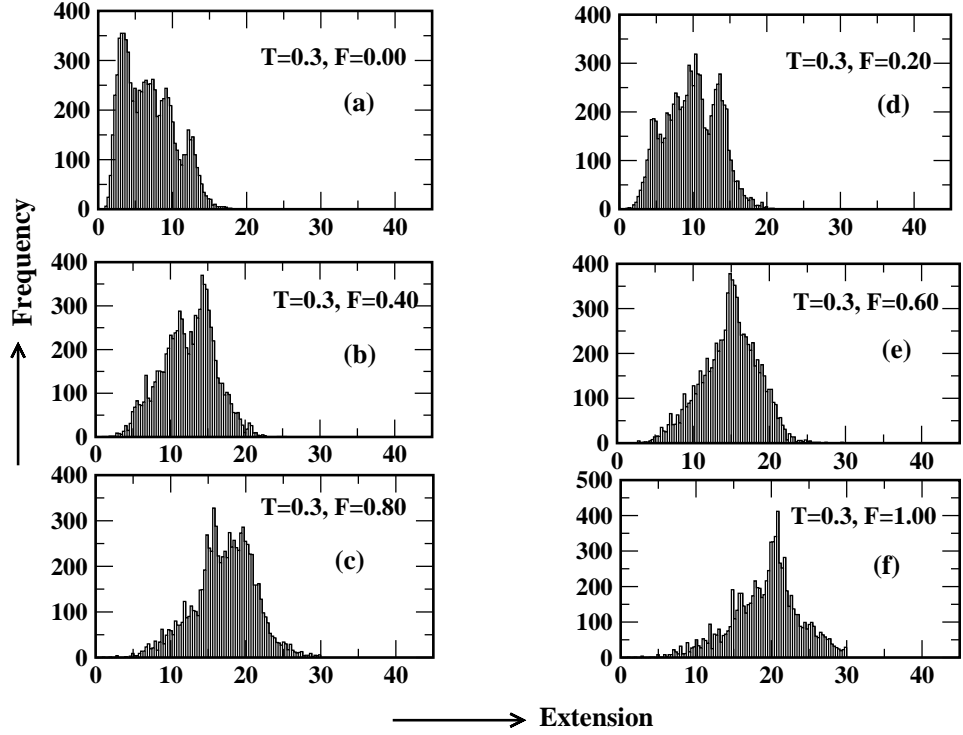}
\caption{Histogram of extension of various realizations at different
forces and fixed temperature ($T = 0.3$).}
\label{fig-4}
\end{figure*}

Molecular crowding is a dynamical phenomenon
involving the appearance, disappearance and movement of voids.
Therefore, the internal structure of the cell changes continuously.
Recently Singh et al \cite{amit} have studied the effect of 
molecular crowding by considering the underlying lattice 
as percolation cluster where each realization of disorder was 
independent of the previous disorder realization.
In presence of molecular crowding agent, Yuan et al. \cite{yuan} 
has performed the unfolding experiment for the first time. 
In order to model the cellular environment, they used dextran
as crowding agents. From this experiment and from the knowledge
accumulated from the structure of cell, it has been revealed that 
the internal structure of the cell changes over the time, but it 
does not change rapidly. Therefore, they consider the average effect
of crowding agents on the unfolding. Hence their results do not
distinguish between molecular confinement and molecular crowding.
Since, the cell structures at 
different time stages remain correlated with the initial (native)
structure, therefore, each realization should also be correlated  
with the initial (native) structure. 
In order to obtain the set of correlated percolation clusters, 
we choose a percolation cluster (native) and generate other clusters 
by changing certain fraction (in this case $10\%$) 
of the occupied site to unoccupied site and vice versa with 
respect to the native. 
This condition leads to a set of correlated 
percolation clusters, where each cluster is correlated with the 
native but may differ with each other with some order of correlation.  
This also ensures that the concentration of the
crowding agent does not change, but the internal structure may change
as happens in the cell. Since we are near to percolation threshold,
every disorder realization constitute the partition function which
is non zero.
This will permit us to study the effect of cellular environment in 
a much more realistic way.

\begin{figure}[b]
\vspace{.5in}
\includegraphics[width=2.5in]{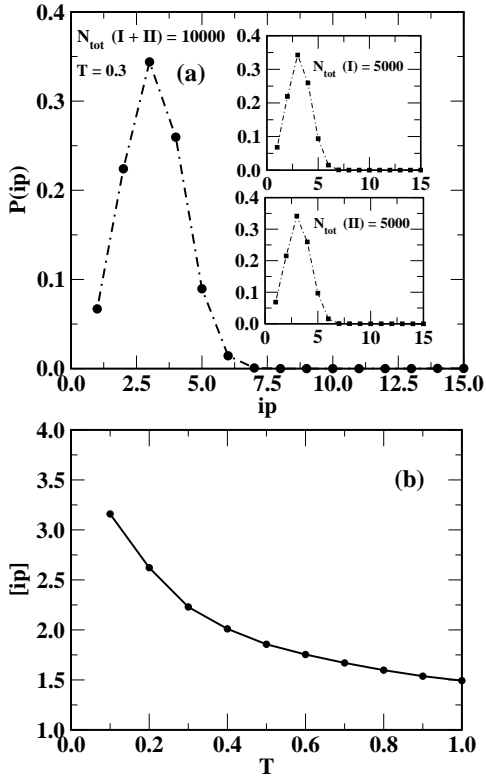}
\caption{(a) Probability of average number of plateaus  at fixed
temperature ($T = 0.3$) for 10000 disorder realizations. Insets show the
same plot for two different 5000 disorder realizations. All these
normalized plot of probability overlaps with each other;
(b) Plot of average number of plateaus ($[ip]$) with temperature for
sample averaged quenched case.
}
\label{fig-5}
\end{figure}

\subsection{Exact Enumeration Method}

We enumerate all possible walks both on square lattice 
with disorder (percolation lattice) and without disorder (pure lattice).
Since, all the single molecule experiments are of few monomers and hence
the observed effect may have finite size effect. In this context  
exact enumeration technique has been proved to be quite successful 
in order to have better understanding of single molecule experiments 
of finite size. As the number of conformation increases with 
$\mu^N$, we restrict our 
calculation for chain length up to 45 steps walk. 
Here $\mu$ is the connectivity 
constant of the lattice.
The partition function of the $i^{th}$ realization of the disorder
configuration may be written as
\begin{equation}
Z^{(i)} = \sum_{(N_p, |x|)} C_N^{(i)} (N_p, | x |)
u^{N_p} \omega^{| x |}.
\end{equation}
Here $C_N^{(i)} (N_p, | x |)$ is
the number of distinct conformations of walks of length $N$ in the $i$-th
realization of the disorder with $N_p$
non-bonded nearest neighbor pairs and whose end-points are a
distance $x$ apart.
$\omega (= \exp[\beta ({\bf F}.\hat{x})])$ is the Boltzmann weight 
for the force, where $\hat{x}$ is the unit vector along 
the {\em x}-axis and $\beta$ is defined as $\frac{1}{k T}$. 
Here, $k$ is the Boltzmann constant and $T$ is the temperature. 
$u (= \exp (-\beta \epsilon_u))$ is the Boltzmann weight 
for nearest neighbor interaction.
 
In the following, we set $\epsilon/k$ = 1 and  focus 
our discussion on force-extension behavior in  CFE and
CDE.  In the present study we consider 10000 realizations  
($N_{tot}$) of the correlated percolation clusters.
In Fig. 2(a), we plot the histogram of $Z$ for all 10000 
realizations. It is surprising to see that each realization which
differs from the native conformation by $10\%$ only give such a 
wide range ($10^3$ to $10^{10}$) of the distribution of $C_N$. The existence of 
long tail in the distribution indicates that the standard deviation will be
quite larger than the mean value.  
The log-normal distribution of the above histogram
of 5000 realization shown in inset of Fig. 2a, give mean value $\approx$
18.2 and standard deviation $\approx$ 3.1 for each set. 
In Fig.2(b), we plot the frequency of $Z$ with rank in the 
log-log scale. From this plot, we find power law 
behavior. In Fig. 2(c), we plot histogram at intermediate 
temperature and force ($T = 0.3$ and $F=0.75$) 
and found that the qualitative behavior (long tail) of 
the histogram and the power law (Fig. 2(d)) remain the same.

\section{Averaging over Disorder}

Following the method developed in ref. \cite{amit}
we denote a sample average over various realizations as 
$[...]$ and thermal averaging as $\langle...\rangle $. 

\subsection{Force-Extension Behavior in CFE}

In CFE, for a given realization of disorder, we calculate the reaction 
coordinate i.e x-component of end-to-end distance from the 
following expression: 

\begin{equation}
\langle x^{(i)}\rangle_Q = \frac{\sum_{(N_p, | x |)} x^{(i)} C_N^{(i)} 
(N_p, | x |) u^{N_p} \omega^{| x |}}{Z^{(i)}} 
\label{qen}
\end{equation}

This may be thought as simple representation of molecular confinement
where disorder sites do not move with time. Therefore, we refer it as 
quenched average of $i$-th realization. 
The sample average over Eq. (\ref{qen}) can be written as
\begin{equation}
[\langle x\rangle_Q] = \sum_i \langle x^{(i)}\rangle_Q/N_{tot} 
\end{equation}
which we call the sample average over quenched disorder. In 
Fig. 3(a) we show some of the representative plots for different 
correlated realizations. The qualitative nature of F-x curve 
almost remain same, but vary for different realizations.
All these plots show the existence of plateaus. It may be 
noted that the number of plateaus may vary from realization to
realization.  The sample averaged reaction coordinate has also been shown 
in Fig. 3(b) at low temperature. 
As pointed above, because of large variation in the partition
functions of different realizations, reaction coordinate also
exhibits the similar variations. In Fig. 4, we plot the histogram
of reaction coordinate for different realizations. With rise of force,
the distribution peak shifts to the right. However, due to the large width
of the histogram, the plateaus seen in individual realization smoothen out
in sample averaging as shown in Fig. 3(b). This does not mean the 
plateau vanishes.
In fact all these plateau exist for different realization. In Fig. 5a, we plot 
the probability distribution of plateaus ($P(ip)$) at $ T=0.3$. The variation of sample 
average of number of plateaus ($[ip]$) with temperature is shown in Fig. 5b. 
From these plots,
it is evident that average number of plateaus decreases with temperature and
system approaches to two-state at high temperature.

\begin{figure}[t]
\includegraphics[width=2.5in]{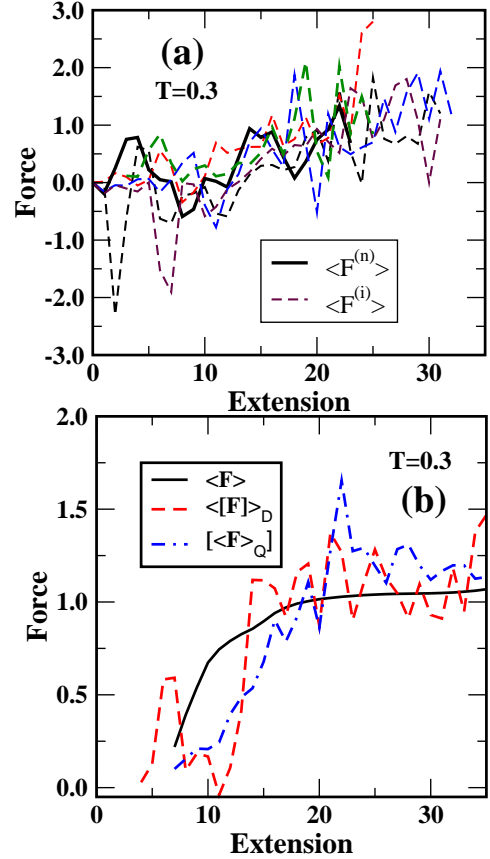}
\caption{Same as Fig. 3, but at CDE for
temperature $T$ = 0.3.
}
\label{fig-6}
\end{figure}

For molecular crowding, we average over finite but large number 
of realizations.  The realization in the limit $N_{tot} \rightarrow
\infty$ is not an experimental requirement. Since, the correlated 
percolation clusters are 
having the structures close to the initial cluster, therefore, 
obtaining all possible percolation clusters are beyond the scope.
Hence we call the averaging over finite correlated clusters 
as an approximate annealed average. It is important to mention here 
that this averaging will vary  on the choice of initial 
cluster as well as on the number of realizations. Moreover because
of long tail in the distribution, two different sets of realizations 
may not give exactly the 
similar F-x curve. However, qualitative nature of the F-x curve 
for two different sets will remain same. 
The reaction coordinate (end-to-end distance) in case of approximate 
annealed averaging has been  defined as 

\begin{equation}
\langle [x]\rangle_D = \frac{\frac{1}{N_{tot}} \sum_i \sum_{(N_p, | x |)} 
x^{(i)} C_N^{(i)}(N_p, | x |)u^{N_p} \omega^{| x |}}{\cal{Z}}
\end{equation}
where $ {\cal{Z}} = \sum_i Z^{(i)}/N_{tot}$ and the summation is over $N_{tot}$
realizations of the disorder. 
Though in the limit $N_{tot} \rightarrow \infty$, annealed average and 
pure give the same result, however, in present case this may not 
be true. Because all the correlated percolation clusters are
having the structures close to the initial cluster and hence all 
possible conformations of disorder are not possible to generate. 
In Fig. 3b, we also 
plot the force extension curve for the approximate annealed case 
which also shows the existence of plateaus / intermediate states. 
At $T=0.3$,
we find two plateaus (Fig. 3b) which is close to the average 
number of plateaus ($[ip]$) obtained by
sample averaged quenched case (Fig. 5b).

\begin{figure*}[t]
\includegraphics[width=5in]{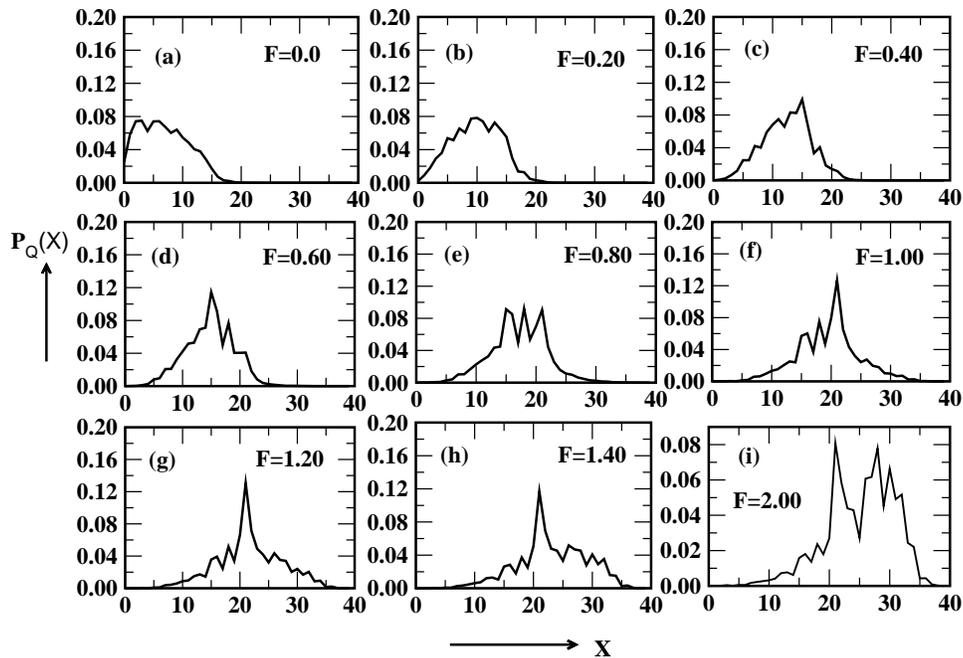}
\caption{Plot of P(x) vs x  for sample averaged quenched disorder case
at different $F$ (fixed temperature $T=0.3$).
}
\label{fig-7}
\end{figure*}

\begin{figure*}[t]
\vspace{.5in}
\includegraphics[width=5in]{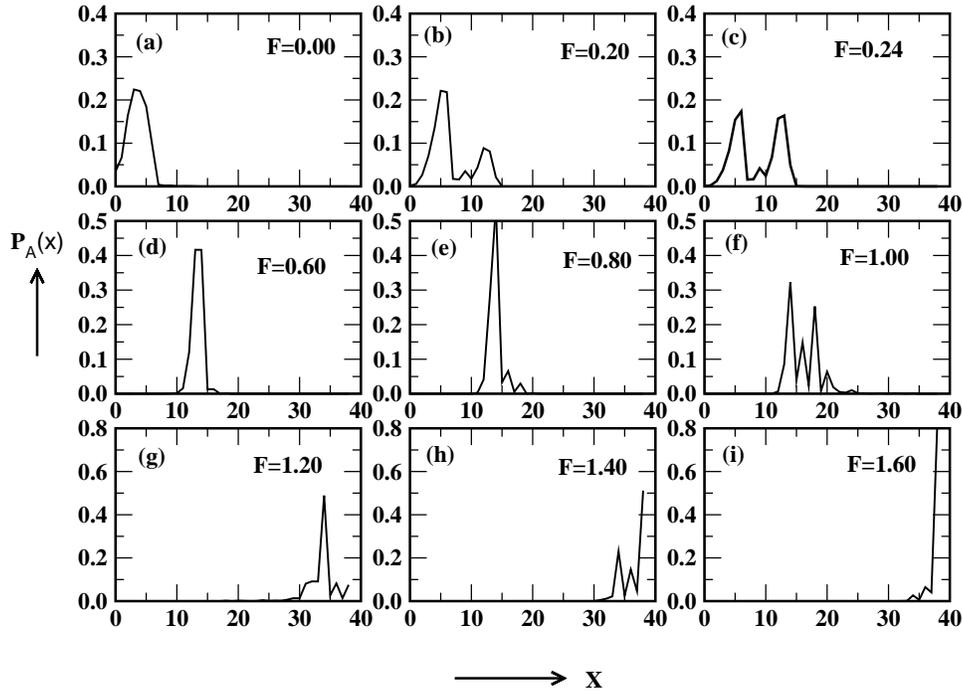}
\caption{Same as Fig. 7, but for the approximate annealed average
case at different $F$ (fixed temperature $T=0.3$). }
\label{fig-8}
\end{figure*}

\subsection{Force-Extension Behavior in CDE}

It is known that atomic force microscopy (AFM) experiments 
work on CDE. When the disorder realizations were independent 
to each other Singh et.al. \cite{amit} have shown strong
oscillations in F-x curve in CDE for approximate annealed 
case. Surprisingly such oscillations were absent for pure 
(in absence of disorder) and sample averaged quenched 
disorder. In present case where each realization is 
correlated, we observe strong oscillations (Fig. 6) in both 
approximate annealed and sample averaged quenched cases.
This may be because in CDE the system probes local ground state 
energy which may be quite close for each disorder realization. 
However in case of independent realizations local ground state 
energy will be different for each realization and hence sample
averaging will smoothen out such oscillations.

\subsection{Probability Distribution}

In Fig 7 and 8, we show the probability distribution curves of reaction 
coordinate for different forces at fixed temperature. This provides 
important information about the changes in the structural processes 
in cellular environment. 
In our calculation we use the following expressions to calculate probability 
distribution for the approximate annealed and sample averaged quenched cases:
\begin{equation}
P_{A}(x)=\frac{\sum_i \sum_{N_P} C_N^{(i)} (N_p, | x |) u^{N_p} 
\omega^{| x |}}{\sum_i Z^{(i)}}
\end{equation}
 and
\begin{equation}
P_{Q}(x)=\frac{1}{N_{tot}}\sum_i [\frac{\sum_{N_P} 
C_N^{(i)} (N_p, | x |) u^{N_p} \omega^{| x |}}{Z^{(i)}}]
\end{equation}

For pure case the probability distribution curve has been studied 
by Singh et. al. \cite{amit}. For the correlated disorder we find 
qualitatively similar behaviour at lower and higher forces. 
At low force, the peak in the 
probability distribution corresponds to the collapsed state and at high 
force it corresponds to the stretched state.  It was found that 
for the pure case at certain force (near the transition point) 
the peak broadens which indicates that the transition is continuous. 
However, for the correlated disorder case (for both approximate annealed  
and sample averaged quenched disorder), one can see the 
emergence and disappearance of  peaks [Fig 7 and 8] for different 
$F$ at $T = 0.3$.
In case of sample averaged quenched case the probability distribution 
profile is broad in nature with many small sharp peaks [Fig. 7]. 
Since entropy associated with sample averaged quenched disorder 
is much higher, therefore, the width of the peak broadens. 
However, for approximate annealed case the width of
the peak remains smaller which is evident from Fig. 8.
With increasing force, we see the height of one  peak increases 
while others decreases.  We find at many different forces the 
height of two peaks (at different position) become equal indicating 
the existence of two states. Unlike pure case, this reflects that 
disorder may change the order of transition and there will be 
emergence of intermediate states which are solely induced by the 
disorder.  

\section{Conclusions}

We have studied the effect of an applied force on globule-coil
transition on the percolation cluster. Simple model presented 
here clearly shows that force induced transition as observed 
in vitro is quite different than in vivo. It is evident from
our model studies that molecular crowding or molecular
confinement induces intermediate states which may be seen 
in the form of plateaus in F-x curve at CFE. It is interesting 
to note that  quenched average over different disorder also show the
existence of intermediate states  similar to approximate annealed case. 
Because of the heterogeneity in the number of
non-bonded nearest neighbor interaction (number may vary 
in between 0 to 2) along the chain and the structure of disorder, 
we see oscillations in F-x 
curve in CDE for both approximate annealed and sample 
averaged quenched cases which were found to be absent for the
uncorrelated disorder \cite{amit}. 

There are many biological events which are rare in nature. 
The occurrence of such event may be because of long-tail 
behavior in the distribution of partition function. Therefore, 
at this moment of time additional numerical
and experimental works are required to understand the 
effect of crowding agents on unfolding processes seen in the
cell.

\section{Acknowledgment}
We thank Department of Science and Technology and University Grants
Commission, New Delhi for financial support.  We acknowledge the 
generous computers support from MPIPKS, Dresden, Germany.

\end{document}